\documentclass{amsart}
\usepackage{hyperref}

\newtheorem{theorem}{Theorem}[section]
\newtheorem{lemma}[theorem]{Lemma}
\newtheorem{corollary}[theorem]{Corollary}
\newtheorem{remark}[theorem]{Remark}
\newtheorem{hypo}[theorem]{Hypothesis {\bf H.}\hspace*{-0.6ex}}

\newcommand{\R}{{\mathbb R}}
\newcommand{\N}{{\mathbb N}}
\newcommand{\Z}{{\mathbb Z}}
\newcommand{\C}{{\mathbb C}}
\newcommand{\M}{{\mathbb M}}

\newcommand{\nn}{\nonumber}
\newcommand{\be}{\begin{equation}}
\newcommand{\ee}{\end{equation}}
\newcommand{\bea}{\begin{eqnarray}}
\newcommand{\eea}{\end{eqnarray}}
\newcommand{\ul}{\underline}
\newcommand{\ti}{\tilde}

\newcommand{\spr}[2]{\langle #1 , #2 \rangle}

\newcommand{\I}{\mathrm{i}}

\newcommand{\re}{\mathrm{Re}}

\newcommand{\lz}{\ell^2(\Z)}
\newcommand{\tl}{\mathrm{TL}}
\newcommand{\tr}{\mathrm{tr}}
\newcommand{\Ker}{\mathrm{Ker}}

\newcommand{\ulz}{\ul{z}}
\newcommand{\hmu}{\hat{\mu}}

\newcommand{\dimuz}{\di_{\ul{\hat{\mu}}}}

\newcommand{\vrc}{\ul{\Xi}_{p_0}}
\newcommand{\hvrc}{\ul{\hat{\Xi}}_{p_0}}

\newcommand{\di}{\mathcal{D}}

\newcommand{\Amap}{\ul{A}_{p_0}}
\newcommand{\amap}{\ul{\alpha}_{p_0}}
\newcommand{\hAmap}{\ul{\hat{A}}_{p_0}}
\newcommand{\hamap}{\ul{\hat{\alpha}}_{p_0}}

\newcommand{\Rg}[1]{R_{2g+2}^{1/2}(#1)}

\newcommand{\eps}{\varepsilon}

\newcommand{\sig}{\sigma}
\newcommand{\lam}{\lambda}
\newcommand{\gam}{\gamma}
\newcommand{\om}{\omega}

\numberwithin{equation}{section}

\begin{document}

\title[Inverse Scattering Transform for the Toda Hierarchy]{Inverse Scattering 
Transform for the Toda Hierarchy with Quasi-Periodic Background}

\author[I. Egorova]{Iryna Egorova}
\address{Kharkiv National University\\ 47,Lenin ave\\ 61164 Kharkiv\\ Ukraine}
\email{\href{mailto:egorova@ilt.kharkov.ua}{egorova@ilt.kharkov.ua}}

\author[J. Michor]{Johanna Michor}
\address{Faculty of Mathematics\\
Nordbergstrasse 15\\ 1090 Wien\\ Austria\\ and International Erwin Schr\"odinger
Institute for Mathematical Physics, Boltzmanngasse 9\\ 1090 Wien\\ Austria}
\email{\href{mailto:Johanna.Michor@esi.ac.at}{Johanna.Michor@esi.ac.at}}
\urladdr{\href{http://www.mat.univie.ac.at/~jmichor/}{http://www.mat.univie.ac.at/\~{}jmichor/}}

\author[G. Teschl]{Gerald Teschl}
\address{Faculty of Mathematics\\
Nordbergstrasse 15\\ 1090 Wien\\ Austria\\ and International Erwin Schr\"odinger
Institute for Mathematical Physics, Boltzmanngasse 9\\ 1090 Wien\\ Austria}
\email{\href{mailto:Gerald.Teschl@univie.ac.at}{Gerald.Teschl@univie.ac.at}}
\urladdr{\href{http://www.mat.univie.ac.at/~gerald/}{http://www.mat.univie.ac.at/\~{}gerald/}}

\thanks{Work supported by the Austrian Science Fund (FWF) under Grant
No.\ P17762 and INTAS Research Network NeCCA 03-51-6637.
}


\begin{abstract}
We provide a rigorous treatment of the inverse scattering transform for
the entire Toda hierarchy in the case of a quasi-periodic finite-gap
background solution. 
\end{abstract}

\maketitle

\section{Introduction}

Since the seminal work of Gardner et al.\ \cite{ggkm} in 1967 the inverse scattering
transform is one of the main tools for solving completely integrable wave equations
and numerous articles have been devoted to this subject since then. However,
with only a few exceptions, only the case of short range perturbations of a
constant background solution has been investigated. The first to consider a
non-constant background seem to be Kuznetsov and A.V. Mikha\u\i lov, \cite{kumi}, who
informally treat the Korteweg-de Vries equation with the Weierstra{\ss} elliptic function
as stationary background solution. Formulas for arbitrary periodic background were
given by Firsova in \cite{firkdv}. However, the discontinuous behavior at the band edges
is not investigated there and neither is the question when the time dependent scattering
data satisfy the hypothesis necessary for the Gel'fand-Levitan-Marchenko theory
(see, e.g., \cite{mar} for the constant background case). Moreover, the normalization
chosen there blows up when the eigenvalues lie at certain positions in the gaps.

Our aim in the present paper is to provide a rigorous treatment of the inverse
scattering transform for the Toda hierarchy in the case of quasi-periodic finite-gap
background solutions (including the case of arbitrary periodic solutions). We include
a discussion of the problems arising from the poles of the Baker-Akhiezer functions
at the Dirichlet eigenvalues (which are absent in the constant background case). 
In the constant background case this was first done for the Toda equation by
Flaschka in \cite{fl2} on an informal level. The first rigorous treatment for the
entire Toda hierarchy was given by one of us in \cite{tist} (see also
\cite{fad}, \cite{ta}, \cite{tjac}). For further results on solutions of the Toda lattice
with nontrivial spatial asymptotics see \cite{bdme2}.

After introducing the Toda hierarchy in Section~\ref{secth}, we will first show that
a solution will stay close to a given background solution in Section~\ref{secivp}.
This result implies that a short range perturbation of a quasi-periodic finite-gap
solution will stay short range for all time and it shows that the time dependent scattering
data satisfy the hypothesis necessary for the Gel'fand-Levitan-Marchenko theory.
This result constitutes the main technical ingredient for the inverse scattering
transform. In Section~\ref{secqp} we review some necessary facts on
quasi-periodic finite-gap solutions and in Section~\ref{secist} we compute
the time dependence of the scattering data and discuss its dynamics. Finally
we establish the conserved quantities of the Toda hierarchy in this context.

\section{The Toda hierarchy}
\label{secth}

In this section we introduce the Toda hierarchy using the standard Lax formalism
(\cite{lax}). We first review some basic facts from \cite{bght} (see also \cite{tjac}).

We will only consider bounded solutions and hence require

\begin{hypo} \label{H t0}
Suppose $a(t)$, $b(t)$ satisfy
\[
a(t) \in \ell^{\infty}(\Z, \R), \qquad b(t) \in \ell^{\infty}(\Z, \R), \qquad
a(n,t) \neq 0, \qquad (n,t) \in \Z \times \R,
\]
and let $t \mapsto (a(t), b(t))$ be differentiable in 
$\ell^{\infty}(\Z) \oplus \ell^{\infty}(\Z)$.
\end{hypo}

\noindent
Associated with $a(t), b(t)$ is a Jacobi operator
\begin{equation}
H(t): \lz  \to  \lz, \qquad f \mapsto \tau(t) f,
\end{equation}
where
\begin{equation}
\tau(t) f(n)= a(n,t) f(n+1) + a(n-1,t) f(n-1) + b(n,t) f(n)
\end{equation}
and $\lz$ denotes the Hilbert space of square summable (complex-valued) sequences
over $\Z$. Moreover, choose constants $c_0=1$, $c_j$, $1\le j \le r$, $c_{r+1}=0$, set
\bea \nn
g_j(n,t) &=& \sum_{\ell=0}^j c_{j-\ell} \spr{\delta_n}{H(t)^\ell \delta_n},\\ \label{todaghsp}
h_j(n,t) &=& 2 a(n,t) \sum_{\ell=0}^j c_{j-\ell}  \spr{\delta_{n+1}}{H(t)^\ell
\delta_n} + c_{j+1},
\eea
and consider the Lax operator
\begin{equation}  \label{btgptdef}
P_{2r+2}(t) = -H(t)^{r+1} + \sum_{j=0}^r ( 2a(t) g_j(t) S^+ -h_j(t)) H(t)^{r-j} +
g_{r+1}(t),
\end{equation}
where $S^\pm f(n) = f(n\pm1)$. Restricting to the two-dimensional nullspace $\Ker(\tau(t)
-z)$, $z\in\C$, of $\tau(t)-z$, we have the following representation of $P_{2r+2}(t)$
\begin{equation} \label{btqptFG}
P_{2r+2}(t)\Big|_{\Ker(\tau(t)-z)} =2a(t) G_r(z,t) S^+ - H_{r+1}(z,t),
\end{equation}
where $G_r(z,n,t)$ and $H_{r+1}(z,n,t)$ are monic polynomials in $z$ of the
type
\bea \nn
G_r(z,n,t) &=& \sum_{j=0}^r z^j g_{r-j}(n,t),\\ \label{fgdef}
H_{r+1}(z,n,t) &=& z^{r+1} + \sum_{j=0}^r z^j h_{r-j}(n,t) - g_{r+1}(n,t).
\eea
A straightforward computation shows that the Lax equation
\begin{equation} \label{laxp}
\frac{d}{dt} H(t) -[P_{2r+2}(t), H(t)]=0, \qquad t\in\R,
\end{equation}
is equivalent to
\bea \nn
\tl_r (a(t), b(t))_1 &=& \dot{a}(t) -a(t) \Big(g_{r+1}^+(t) -
g_{r+1}(t) \Big)=0,\\ \label{tlrabo}
\tl_r (a(t), b(t))_2 &=& \dot{b}(t) - \Big(h_{r+1}(t) -h_{r+1}^-(t) \Big)=0,
\eea
where the dot denotes a derivative with respect to $t$ and $f^\pm(n)=f(n\pm 1)$.
Varying $r\in \N_0$ yields the Toda hierarchy 
$\tl_r(a,b) =(\tl_r (a,b)_1, \tl_r (a,b)_2) =0$.

Finally, we recall that the Lax equation (\ref{laxp}) implies existence of
a unitary propagator $U_r(t,s)$ such that the family of operators
$H(t)$, $t\in\R$, are unitarily equivalent, $H(t) = U_r(t,s) H(s) U_r(s,t)$.

\section{The initial value problem}
\label{secivp}

First of all we recall the basic existence and uniqueness theorem for the Toda hierarchy
(see, e.g., \cite{tist}, \cite{ttkm}, or \cite[Section~12.2]{tjac}).

\begin{theorem} \label{thmexistandunique}
Suppose $(a_0,b_0) \in M = \ell^\infty(\Z) \oplus \ell^\infty(\Z)$. Then there exists a unique
integral curve $t \mapsto (a(t),b(t))$ in $C^\infty(\R,M)$ of the Toda equations, that is,
$\tl_r(a(t),b(t))=0$, such that $(a(0),b(0)) = (a_0,b_0)$.
\end{theorem}

\noindent
In \cite{tist} it was shown that solutions which are asymptotically close to the constant
solution at the initial time stay close for all time. Our first aim is to extend this
result to include perturbations of quasi-periodic finite-gap solutions. In fact,
we will even be a bit more general.

\begin{lemma} \label{lemtsr}
Suppose $a(n,t)$, $b(n,t)$ and $\bar a(n,t)$, $\bar b(n,t)$ are two arbitrary 
bounded solutions of the Toda system satisfying
(\ref{H t2}) for one $t_0 \in \R$, then (\ref{H t2}) holds for all 
$t \in \R$, that is, 
\be \label{H t2}
\sum_{n \in \Z} w(n) \Big(|a(n,t) - \bar a(n,t)| + |b(n,t) - \bar b(n,t)| \Big) 
< \infty,
\ee
where $w(n)>0$.
\end{lemma}

\begin{proof}
Without loss of generality we assume that $t_0 = 0$. Consider the  
expression
\[
\| (a(t),b(t))\|_* = 
\sum_{n \in \Z}w(n) \big(|a(n,t) - \bar a(n,t)| 
+ |b(n,t) - \bar b(n,t)| \big) 
\]
which remains finite at least for small $t$ since there is a local solution 
of the Toda system with respect to $\|.\|_*$. We claim that
the following estimate holds
\begin{align} \nn 
\sum_{n \in \Z}w(n) |g_r(n,t) - \bar g_r(n,t)| 
&\leq C_r \| (a(t),b(t))\|_*,  \\ \label{H t3}
\sum_{n \in \Z}w(n) |h_r(n,t) - \bar h_r(n,t)| 
&\leq C_r \| (a(t),b(t))\|_*,
\end{align}
where $C_r=C\big(\|H(0)\|,\|\bar H(0)\|\big)$ is a positive constant due to
H.\ref{H t0}.
Let us prove (\ref{H t3}) by induction on $r$. It suffices to consider the case 
where $c_j=0$, $1\leq j \leq r$, since all involved sums are finite. 
In this case \cite[Lemma~6.4]{tjac} shows that 
$g_j(n,t)$, $h_j(n,t)$ can be recursively computed from 
$g_0(n,t)=1$, $h_0(n,t)=0$ via
\begin{align} \nn
g_{j+1}(n,t) &= \frac{1}{2} \big(h_j(n,t) + h_j(n-1,t)\big)
+ b(n,t) g_j(n,t), \\ \nonumber
h_{j+1}(n,t)&= 2 a(n,t)^2 \sum_{l=0}^j g_{j-l}(n,t)g_l(n+1,t)
- \frac{1}{2} \sum_{l=0}^j h_{j-l}(n,t)h_l(n,t).  
\end{align}
Hence  
\[
g_1(n,t) - \bar g_1(n,t) = b(n,t) - \bar b(n,t), \quad 
h_1(n,t) - \bar h_1(n,t) = 2 (a(n,t)^2 - \bar a(n,t)^2),
\]
and 
\begin{align}  \nonumber
&g_{r+1}(n,t) - \bar g_{r+1}(n,t)  
= \big( b(n,t) - \bar b(n,t)\big) g_r(n,t) + 
\bar b(n,t) \big( g_r(n,t) - \bar g_r(n,t)\big) \\ \nonumber
& \quad + \frac{1}{2} \big(h_r(n,t) - \bar h_r(n,t) + h_r(n-1,t) -
\bar h_r(n-1,t) \big), \\ \nonumber
& h_{r+1}(n,t) - \bar h_{r+1}(n,t) = \frac{1}{2} \sum_{l=0}^r 
\big(h_{r-l}(n,t)h_l(n,t) - \bar h_{r-l}(n,t)\bar h_l(n,t)\big)\\ \nonumber
& \quad + 2 a(n,t)^2 \sum_{l=0}^r g_{r-l}(n,t)g_l(n+1,t)
- 2 \bar a(n,t)^2 \sum_{l=0}^r \bar g_{r-l}(n,t)\bar g_l(n+1,t).
\end{align}
Since both sequences $a$, $b$ and $\bar a$, $\bar b$
are solutions of the Toda system, (\ref{tlrabo}) yields
\begin{align*}  
&\dot a(t) - \dot{\bar a}(t) = 
a(t)\big(g_{r+1}^+(t) - g_{r+1}(t)\big) -
\bar a(t)\big(\bar g_{r+1}^+(t) - g_{r+1}(t)\big) 
\\ \nn
&\quad =(a(t)-\bar a(t))\big(\bar g_{r+1}^+(t) - \bar g_{r+1}(t)\big) 
- a(t)\big(\bar g_{r+1}^+(t)-g_{r+1}^+(t) 
+ g_{r+1}(t) - \bar g_{r+1}(t)\big) 
\end{align*}
and thus
\begin{align} \nonumber
&|a(n,t) - \bar a(n,t)| \leq  \big|a(n,0) -\bar a(n,0)\big| +
C_1 \int_0^t \big|a(n,s) -\bar a(n,s)\big| ds \\ \nonumber
& \quad + \|H(0)\| \int_0^t \big(|g_{r+1}(n+1,s) - \bar g_{r+1}(n+1,s)| 
+ |g_{r+1}(n,s) - \bar g_{r+1}(n,s)|\big)ds,
\\ \nonumber
&|b(n,t) - \bar b(n,t)| \leq |b(n,0) - \bar b(n,0)|  \\ \nonumber 
& \quad + \int_0^t
(|h_{r+1}(n,s) - \bar h_{r+1}(n,s)| + 
|h_{r+1}(n-1,s) - \bar h_{r+1}(n-1,s)|)ds.
\end{align}
By (\ref{H t3}),
$
\| (a(t),b(t))\|_* \leq   \| (a(0),b(0))\|_* + 
C \int_0^t \| (a(s),b(s))\|_*ds,
$
where $C = (C_1 + 2\|H(0)\| +2) C_{r+1}$.
Finally, applying Gronwall's inequality
\[
\| (a(t),b(t))\|_* \leq \| (a(0),b(0))\|_*  
\exp\big(C t\big)
\]
finishes the proof.
\end{proof}

\section{Quasi-periodic finite-gap solutions}
\label{secqp}

As a preparation for our next section we first need to recall some facts on
quasi-periodic finite-gap solutions (again see \cite{bght} or \cite{tjac}).

Let $\M$ be the Riemann surface associated with the following function
\begin{equation}
\Rg{z}= -\prod_{j=0}^{2g+1} \sqrt{z-E_j}, \qquad
E_0 < E_1 < \cdots < E_{2g+1},
\end{equation}
where $g\in \N$ and $\sqrt{.}$ is the standard root with branch cut along $(-\infty,0)$. 
$\M$ is a compact, hyperelliptic Riemann surface of genus $g$.
A point on $\M$ is denoted by 
$p = (z, \pm \Rg{z}) = (z, \pm)$, $z \in \C$, or $p = \infty_{\pm}$, and
the projection onto $\C \cup \{\infty\}$ by $\pi(p) = z$. 

Now pick $g$ numbers (the Dirichlet eigenvalues)
\be
(\hat{\mu}_j)_{j=1}^g = (\mu_j, \sigma_j)_{j=1}^g
\ee
whose projections lie in the spectral gaps, that is, $\mu_j\in[E_{2j-1},E_{2j}]$.
Associated with these numbers is the divisor $\dimuz$ which
is one at the points $\hat{\mu}_j$  and zero else. Using this divisor we
introduce
\begin{align} \nn
\ulz(p,n,t) &= \hAmap(p) - \hamap(\dimuz) - n\ul{\hat A}_{\infty_-}(\infty_+)
+ t\ul{U}_s - \hvrc \in \C^g, \\
\ulz(n,t) &= \ulz(\infty_+,n,t),
\end{align}
where $\vrc$ is the vector of Riemann constants, 
$\ul{U}_s$ the $b$-periods of the Abelian differential
$\Omega_s$ defined below, and $\Amap$ ($\amap$)
is Abel's map (for divisors). The hat indicates that we
regard it as a (single-valued) map from $\hat{\M}$ (the fundamental polygon
associated with $\M$) to $\C^g$.
We recall that the function $\theta(\ulz(p,n,t))$ has precisely $g$ zeros
$\hmu_j(n,t)$ (with $\hmu_j(0,0)=\hmu_j$), where $\theta(\ul{z})$ is the
Riemann theta function of $\M$.

Then our solution is given by
\begin{align} \nn
a_q(n,t)^2 &= \ti{a}^2 \frac{\theta(\ulz(n+1,t)) \theta(\ulz(n-1,t))}{\theta(
\ulz(n,t))^2},\\ \label{imfab}
b_q(n,t) &= \tilde{b} + \sum_{j=1}^g c_j(g)
\frac{\partial}{\partial w_j} \ln\Big(\frac{\theta(\ul{w} +
\ulz(n,t)) }{\theta(\ul{w} + \ulz(n-1,t))}\Big) \Big|_{\ul{w}=0}.
\end{align}
The constants $\ti{a}$, $\tilde{b}$, $c_j(g)$ depend only on the Riemann surface
(see \cite[Section~9.2]{tjac}).

Introduce
\begin{align} \nn
\phi_q(p,n,t) &= C(n,t) \frac{\theta (\ulz(p,n+1,t))}{\theta (\ulz(p,n,t))}
\exp \Big( \int_{p_0}^p \om_{\infty_+,\infty_-} \Big),\\ \label{BAfthetarep} \quad
\psi_q(p,n,t) &= C(n,0,t) \frac{\theta (\ulz(p,n,t))}{\theta(\ulz (p,0,0))}
\exp \Big( n \int_{p_0}^p \om_{\infty_+,\infty_-} + t\int_{p_0}^p \Omega_s
\Big),
\end{align}
where $C(n,t)$, $C(n,0,t)$ are real-valued,
\begin{equation}
C(n,t)^2 = \frac{\theta(\ulz(n-1,t))}{\theta(\ulz(n+1,t))}, \qquad
C(n,0,t)^2 = \frac{ \theta(\ulz(0,0)) \theta(\ulz(-1,0))}
{\theta (\ulz (n,t))\theta (\ulz (n-1,t))},
\end{equation}
and the sign of $C(n,t)$ is opposite to that of $a_q(n,t)$.
$\om_{\infty_+,\infty_-}$ is the Abelian differential of the third kind with poles
at $\infty_+$ respectively $\infty_-$ and $\Omega_s$ is an Abelian differential
of the second kind with poles at $\infty_+$ respectively $\infty_-$ whose
Laurent expansion is given by the coefficients $(j+1)\hat{c}_{s-j}$ associated with
$\hat\tl_s$ (see \cite[Sections~13.1, 13.2]{tjac}). Then
\bea  \nonumber
\tau_q(t) \psi_q(p,n,t) &=& \pi(p) \psi_q(p,n,t), \\ \nonumber
\frac{d}{dt} \psi_q(p,n,t) &=& 2 a_q(n,t) \hat G_s(p,n,t) 
\psi_q(p,n+1,t) - \hat H_{s+1}(p,n,t) \psi_q(p,n,t) \\ \label{d/dt psi 2}
&=& \hat P_{q,2s+2}(t) \psi_q(p,n,t)(n).
\eea

It is well known that the spectrum of $H_q(t)$ is time independent and
consists of $g+1$ bands
\begin{equation}
\sig(H_q) = \bigcup_{j=0}^g [E_{2j},E_{2j+1}].
\end{equation}
For further information and proofs we refer to \cite[Chapter~9]{tjac}.

Next let us renormalize the Baker-Akhiezer function
\[ 
\tilde \psi_q(p,n,t) = \frac{\psi_q(p,n,t)}{\psi_q(p,0,t)}
\]
such that $\tilde \psi_q(p,0,t)=1$ and let us abbreviate
\be \label{defalpha}
\exp\big(\alpha_s(p,t)\big)= \psi_q(p,0,t) =
C(0,0,t) \frac{\theta (\ulz(p,0,t))}{\theta(\ulz (p,0,0))}
\exp \Big( t\int_{p_0}^p \Omega_s \Big).
\ee
One can show (\cite{tjac}, (13.49))
\begin{align} \nn
\exp\big(\alpha_s(p,t)\big) &=
\exp\bigg( \int_0^t \big(2 a_q(0,x) \hat G_s(p,0,x) \phi_q(p,0,x) 
- \hat H_{s+1}(p,0,x) \big)dx \bigg)\\ \label{alpha G}
&=  \sqrt{\frac{G_g(p,0,t)}{G_g(p,0,0)}} \exp\bigg( R^{1/2}_{2g+2}(p)
\int_0^t \frac{\hat G_s(p,0,x)}{G_g(p,0,x)}dx\bigg),
\end{align}
where
\be
G_g(z,n,t)= \prod_{j=1}^g (z-\mu_j(n,t)).
\ee
However, let us emphasize that the integrals in (\ref{alpha G}) are only well-defined
until we hit the first pole of the integrand. After that time they have to be interpreted
as principal values (compare also Remark~\ref{remtev}~(i)), respectively
(\ref{defalpha}) has to be used.

In addition, observe
\begin{align} \nn
\exp \big(\alpha_{s,+}(z,t) + \alpha_{s,-}(z,t)\big) &=
\frac{G_g(z,0,t)}{G_g(z,0,0)},\\ \label{pmals}
\exp \big(\alpha_{s,+}(z,t) - \alpha_{s,-}(z,t)\big) &=
\exp\bigg(2 R^{1/2}_{2g+2}(z) \int_0^t \frac{\hat G_s(z,0,x)}{G_g(z,0,x)}dx\bigg).
\end{align}
For $\lambda \in \sigma(H_q)$ we have
\be \label{alpha G 2}
\overline{\alpha_{s, \pm}(\lambda,t)} =  \alpha_{s, \mp}(\lambda,t),
\ee 
but note that $\alpha_{s, \mp}(\lambda,t) \neq \mp \alpha_{s, \pm}(\lambda,t)$
unless $g=0$.

\section{Inverse scattering transform}
\label{secist}

Let $a(n,t)$, $b(n,t)$ be a solution of the Toda system satisfying
\be \label{hypo}
\sum_{n \in \Z} |n| \Big(|a(n,t) - a_q(n,t)| + |b(n,t) - b_q(n,t)| \Big) 
< \infty 
\ee
for one (and hence for any) $t_0 \in \R$. 
In \cite{emtqps} (see also \cite{voyu}) we develop scattering theory for the Jacobi operator $H$ 
associated with $a(n)$, $b(n)$. 
Jost solutions, transmission and reflection coefficients depend now on an 
additional parameter $t \in \R$. 
More precisely, 
\be
\sigma(H(t)) \equiv \sigma(H), \quad 
\sigma_{ess}(H)=\sigma(H_q), \quad 
\sigma_p(H)=\{\rho_j\}_{j=1}^q \subseteq \R \backslash \sigma(H_q),
\ee
where $q \in \N$ is finite. The Jost solutions $\psi_{\pm}(z,n,t)$ are
normalized such that 
\[
\tilde\psi_\pm(z,n,t)= \tilde\psi_{q,\pm}(z,n,t)\,(1 + o(1)) \quad
\mbox{as } n\to\pm\infty.
\]

Transmission $T(\lambda,t)$ and reflection $R_{\pm}(\lambda,t)$ 
coefficients are defined via the scattering relations
\be
T(\lambda,t) \tilde\psi_{\pm}(\lambda,n,t)= \overline{\tilde\psi_{\mp}(\lambda,n,t)} + 
R_{\mp}(\lambda,t) \tilde\psi_{\mp}(\lambda,n,t), \qquad \lambda \in \sigma(H_q),
\ee 
and the norming constants $\gamma_{\pm, j}(t)$ corresponding to $\rho_j \in \sigma_p(H)$
are given by
\be
\gamma_{\pm, j}(t)^{-1}=\sum_{n \in \Z}|\tilde \psi_\pm(\rho_j, n,t)|^2.
\ee
To avoid the poles of the Baker-Akhiezer function, we will assume that none of the
eigenvalues $\rho_j$ coincides with a Dirichlet eigenvalue $\mu_k(0,0)$. This
can be done without loss of generality by shifting the initial time $t_0=0$ if
necessary. 

\begin{remark}
Due to this assumption there is no need to remove these poles for the definition
of $\gamma_{\pm, j}$, as we did in \cite{emtqps}. Since the
Dirichlet eigenvalues rotate in their gap, the factor needed to remove the poles
would only unnecessarily complicate the time evolution of the norming constants.
Moreover, these factors would eventually cancel in the Gel'fand-Levitan-Marchenko
equation, which is the only interesting object from the inverse spectral point of
view in the first place. See Remark~\ref{remtev} below.
\end{remark}

\begin{lemma} \label{th: sol H t}
Let $(a(t),b(t))$ be a solution of the Toda hierarchy such that (\ref{hypo}) holds.
The functions 
\be
\psi_{\pm}(z,n,t)=\exp(\alpha_{s, \pm}(z,t)) \tilde\psi_{\pm}(z,n,t) 
\ee
satisfy
\be \label{H t 1}
H(t) \psi_\pm(z,n,t) = z \psi_\pm(z,n,t), \qquad
\frac{d}{dt} \psi_\pm(z,n,t) = \hat P_{2s+2}(t) \psi_\pm(z,n,t).
\ee
\end{lemma}

\begin{proof}
We proceed as in \cite[Theorem 3.2]{tist}. 
The Jost solutions $\psi_{\pm}(z,n,t)$ are continuously differentiable with respect to
$t$ by the same arguments as for $z$ (compare \cite[Theorem 4.2]{emtqps})
and the derivatives are equal to the derivatives of the Baker-Akhiezer functions
as $n \rightarrow \pm \infty$.

For $z \in \rho(H)$, the solution $u_{\pm}(z,n,t)$ of (\ref{H t 1}) with initial condition 
$\psi_{\pm}(z,n,0) \in \ell_{\pm}^2(\Z)$ remains square summable near $\pm \infty$
for all $t \in \R$ (see \cite{ttkm} or \cite[Lemma 12.16]{tjac}), that is,
$u_{\pm}(z,n,t) = C_\pm(t) \psi_{\pm}(z,n,t)$. Letting $n\to\pm\infty$ we
see $C_\pm(t)=1$. The general result for all $z \in \C$ now follows from continuity.
\end{proof}

\noindent
This implies

\begin{theorem} \label{thmtdscat}
Let $(a(t),b(t))$ be a solution of the Toda hierarchy such that (\ref{hypo}) holds.
The time evolution for the scattering data is given by
\bea \nn
T(\lambda, t) &=& T(\lambda, 0), \\ \nn
R_{\pm}(\lambda, t) &=& R_{\pm}(\lambda, 0) 
\exp(\pm(\alpha_{s,+}(z,t)-\alpha_{s,-}(z,t))), \\  
\gamma_{\pm, j}(t) &=& \gamma_{\pm, j}(0) \exp(2 \alpha_{s,\pm}(\rho_j,t)), 
\qquad 1 \leq j \leq q.
\eea
\end{theorem}

\begin{proof}
Since $\overline{\alpha_{s, \pm}(\lambda,t)}= \alpha_{s, \mp}(\lambda,t)$ 
by (\ref{alpha G 2}), we have 
\bea \nonumber
T(\lambda, t) &=& \frac{W(\tilde\psi_-(\lambda,t), \overline{\tilde\psi_-(\lambda,t)})}
{W(\tilde\psi_-(\lambda,t), \tilde\psi_+(\lambda,t))} \\ \nonumber
&=& 
\frac{\exp(\alpha_{s, -}(\lambda,t))\exp(\alpha_{s, +}(\lambda,t))}
{\exp(\alpha_{s, -}(\lambda,t)) \overline{\exp(\alpha_{s, -}(\lambda,t))}}
\frac{W(\psi_-(\lambda,t), \overline{\psi_-(\lambda,t)})}
{W(\psi_-(\lambda,t), \psi_+(\lambda,t))}  = T(\lambda, 0). 
\eea
Here we used that the Wronskian 
$W_n(f,g)=a(n)(f(n)g(n+1)-f(n+1)g(n))$
of two solutions satisfying (\ref{H t 1}) 
does not depend on $n$ or $t$ (see \cite{ttkm}, \cite[Lemma 12.15]{tjac}). Similarly,
\bea \nonumber
R_{\pm}(\lambda, t) &=&
\frac{\exp(\alpha_{s, \pm}(\lambda,t))\exp(\alpha_{s, \mp}(\lambda,t))}
{\exp(\alpha_{s, \mp}(\lambda,t))\exp(\overline{\alpha_{s, \pm}(\lambda,t)})}
\frac{W(\psi_{\mp}(\lambda,t), \overline{\psi_{\pm}(\lambda,t)})}
{W(\psi_{\pm}(\lambda,t), \psi_{\mp}(\lambda,t))} \\ \nn
&=& \exp(\alpha_{s,\pm}(z,t)-\alpha_{s,\mp}(z,t)) R_{\pm}(\lambda, 0).
\eea
The time dependence of $\gamma_{\pm, j}(t)$ follows from
$\|\psi_\pm(\rho_j,.,t)\|= \|\hat U_s(t,0)\psi_\pm(\rho_j,.,0)\|= \|\psi_\pm(\rho_j,.,0)\|$.
\end{proof}

\noindent
A few remarks concerning this theorem are in order:

\begin{remark} \label{remtev}
(i) First of all note that the time dependence of $R_{\pm}(\lambda,t)$ is {\em not}
continuous at the band edges (except the lowest and highest).
In fact, the integrand in (\ref{pmals}) has a pole whenever $\mu_j(0,t)$
hits a band edge. Suppose that $\mu_j(0,t)$ gets close to
$E_{2j}$ at $t=t_1$, then (assuming $G_s(E_{2j},0,t_1)\ne 0$ for simplicity)
\[
\mu_j(0,t) = E_{2j} + \nu_j^2\, (t-t_1)^2 + O(t-t_1)^3, \quad
\nu_j= \frac{G_s(E_{2j},0,t_1) \prod_{k\neq 2j}\sqrt{E_{2j}-E_k}}
{\prod_{l\neq j}(\mu_j(0,t_1) - \mu_l(0,t_1))}.
\]
As long as $t<t_1$ we have
\[
\lim_{\lambda\downarrow E_{2j}} (\alpha_{s,+}(\lambda,t)-\alpha_{s,-}(\lambda,t)) = 0.
\]
But at $t=t_1$ we obtain (setting $\lambda=E_{2j}+\eps^2$ and using
$\nu_j=(-1)^{g-j} \I|\nu_j|$)
\begin{align} \nonumber
& \lim_{\lambda\downarrow E_{2j}} (\alpha_{s,+}(\lambda,t)-\alpha_{s,-}(\lambda,t)) =
\lim_{\lambda\downarrow E_{2j}} 2 R^{1/2}_{2g+2}(\lambda) \int_0^{t_1}
\frac{\hat G_s(\lambda,0,x)}{G_g(\lambda,0,x)}dx\\ \nonumber
& = \lim_{\eps\downarrow 0} 2 \int_0^{t_1}
\frac{\nu_j \eps}{\eps^2 - \nu_j^2\, (x-t_1)^2} dx
 = 2(-1)^{g-j}\I \lim_{\eps\downarrow 0} \arctan(\frac{t_1 |\nu_1|}{\eps}) = (-1)^{g-j} \I \pi.
\end{align}
This is in agreement with the conditions
\be
\begin{array}{l@{\qquad}l}
\lim\limits_{\lambda \rightarrow E_l} R^{1/2}_{2r+2}(\lambda) 
\frac{R_{\pm}(\lambda,t) + 1}{T(\lambda)} = 0,
&  E_l \neq \mu_j(0,t) \\
\lim\limits_{\lambda \rightarrow E_l} R^{1/2}_{2r+2}(\lambda) 
\frac{R_{\pm}(\lambda,t) - 1}{T(\lambda)} = 0,
&   E_l = \mu_j(0,t)
\end{array}, 
\ee
found in \cite{emtqps}.

(ii) The norming constants $\gam_{\pm,j}(t)$ will vanish whenever $(\rho_j,\pm)=
\hat\mu_k(0,t)$. This is important, since the norming constants must cancel
the pole of $\tilde\psi_{q,\pm}(\rho_j,n,t)$ at these points such that the
Gel'fand-Levitan-Marchenko equation remains well-defined.

In fact, these poles come from the normalization $\tilde\psi_{q,\pm}(z,0,t)=1$
and this normalization is natural in the constant background case since it
renders $\tilde\psi_{q,\pm}(z,n,t)$ time-independent. In the (quasi-)periodic
background case this is no longer true and it seems more natural
to allow a time-dependent normalization and replace $\tilde\psi_{q,\pm}(z,n,t)$
by the time-dependent Baker-Akhiezer functions $\psi_{q,\pm}(z,n,t)$ rendering
the norming constants and the reflection coefficient time-independent.

(iii) Finally let us discuss the connection with the different normalizations
for $\gamma_{\pm, j}$ used in \cite{emtqps} respectively \cite{firkdv}.

In \cite{emtqps} we have multiplied $\psi_{q,\pm}(\lambda,n,t)$ by a factor to
cancel the poles at the Dirichlet eigenvalues. However, since the Dirichlet
eigenvalues change sheets during time evolution, this normalization renders
$\hat\psi_{q,\pm}(\lambda,n,t)$ discontinuous with respect to $t$ and
complicates the time evolution of the corresponding norming constants:
\begin{align}  
\hat\gamma_{\pm, j}(t) &= \hat\gamma_{\pm, j}(0) 
\exp\bigg( \pm 2 R^{1/2}_{2g+2}(\rho_j) \int_0^t 
\frac{\hat G_s(\rho_j,0,x)}{G_g(\rho_j,0,x)}dx \\ \nn
&\quad \pm \sum_{l=1}^g \int_0^t \frac{2 \hat G_s(\mu_l(x),0,x) R^{1/2}_{2g+2}(\mu_l(x))}
{(\rho_j - \mu_l(x))\prod_{k\neq l}(\mu_l(x) - \mu_k(x))}dx \bigg).
\end{align}
This clearly justifies our decision to use the freedom of choosing the initial time to avoid the poles in
the first place.

If $H_q$ is periodic with period $N$, the normalization chosen in \cite{firkdv} is to divide
$\psi_{q,\pm}(\lambda,n,t)$ by the square root of $\frac{1}{N} \sum_{n=1}^{N} \psi_{q, +}(z, n) \psi_{q, -}(z, n)$.
To get a well-defined real-valued result this object should be positive. However, since one can compute (see \cite{emtqps})
\be
\sum_{n=1}^{N} \psi_{q, +}(z, n) \psi_{q, -}(z, n) = N \prod_{j=1}^{N-1} \frac{z-\lambda_j}{z-\mu_j},
\ee
where $\lam_j\in[E_{2j-1},E_{2j}]$ are the zeroes of $\omega_{\infty_-,\infty_+}$, this
factor can be negative or even zero. While this choice does give a simple time
dependence 
\be
\tilde\gamma_{\pm, j}(t) = \tilde\gamma_{\pm, j}(0) \exp(\alpha_{s,\pm}(\rho_j,t) - \alpha_{s,\mp}(\rho_j,t)),
\ee
it does not cancel the poles at the Dirichlet eigenvalues (it just renders poles into square
root singularities) plus it introduces additional singularities at $\lambda_j$. In particular,
it is unsuitable if one of the eigenvalues coincides with some $\lambda_j$.
\end{remark}

\noindent
In summary, since Lemma~\ref{lemtsr} ensures that (\ref{hypo}) remains
valid for all $t$ once it holds for the initial condition, we can compute
$R_\pm(\lambda,0)$ and $\gamma_{\pm,j}(0)$ from $(a(n,0), b(n,0))$ and then
solve the Gel'fand-Levitan-Marchenko (GLM) equation to obtain $(a(n,t), b(n,t))$ as in
\cite{emtqps}. In this respect the most important ingredient is the time
dependence of the kernel of the GLM equation which follows from
Theorem~\ref{thmtdscat}.

\begin{theorem}
The time dependence of the kernel of the Gel'fand-Levitan-Marchenko equation is
given by
\bea \nn
F^\pm(m,n,t) &=&
\frac{1}{\pi} \re\int_{\lambda\in\sig(H_q)}
R_\pm(\lambda,0) \psi_{q, \pm}(\lambda, m,t) \psi_{q, \pm}(\lambda, n, t) 
\frac{G_g(\lambda,0,0)}{R^{1/2}_{2g+2}(\lambda)}  d\lambda\\
&&{} + \sum_{j=1}^q \gamma_{\pm,j}(0) \psi_{q, \pm}(\rho_j, m,t) \psi_{q, \pm}(\rho_j, n,t).
\eea
\end{theorem}

\noindent
Finally note that since the transmission coefficient is time independent we obtain

\begin{corollary}
Let $(a(t),b(t))$ be a solution of the Toda hierarchy such that (\ref{hypo}) holds.
Then the quantities
\be
A = \prod_{j=- \infty}^\infty \frac{a(j,t)}{a_q(j,t)}
\ee
and $\tau_j= \tr(H^j(t) - H_q(t)^j)$, that is,
\begin{align} \nn
\tau_1 &= \sum_{n\in\Z} b(n,t) - b_q(n,t)\\ \nn
\tau_2 &= \sum_{n\in\Z} 2( a(n,t)^2 - a_q(n,t)^2) + (b(n,t)^2 - b_q(n,t)^2)
\end{align}
etc., are conserved quantities.
\end{corollary}

\begin{proof}
This follows from the fact that the transmission coefficient
is the perturbation determinant (in the sense of Krein) of the pair $H_q$ and $H$
(see \cite{mtqptr}).
\end{proof}

\end{document}